\documentclass[pre,onecolumn,tightenlines,12pt]{revtex4}
\usepackage{graphicx}
\usepackage{multirow}
\usepackage{CJKutf8}
\usepackage{verbatim}
\usepackage{longtable}
\begin{document}

\title{Random matrices applications to soft spectra}

\author{Rongrong Xie$^{1}$, Weibing Deng$^{1, \dagger}$, and Mauricio P. Pato$^{2, \dagger}$\footnotetext[2]{Authors to whom any 
correspondence should be addressed: wdeng@mail.ccnu.edu.cn and mpato@if.usp.br}}

\affiliation{$^{1}$ Key Laboratory of Quark and Lepton Physics (MOE) and 
Institute of Particle Physics, Central China Normal University, 
Wuhan 430079, China \\
$^{2}$Inst\'{\i}tuto de F\'{\i}sica, Universidade de S\~{a}o Paulo\\
Caixa Postal 66318, 05314-970 S\~{a}o Paulo, S.P., Brazil}

\begin{abstract}
It recently has been found that methods of the statistical theories of spectra can be a useful
tool in the analysis of spectra far from levels of Hamiltonian systems. Several examples originate from areas, 
such as quantitative linguistics and polymers. The purpose of the present study is to deepen
this kind of approach by performing a more 
comprehensive spectral analysis that measures both the local and long-range statistics.  
We have found that, as a common feature, spectra of this kind can exhibit a situation in which
local statistics are relatively quenched while the long range ones show large fluctuations. By combining extensions of the standard 
Random Matrix Theory (RMT) and considering long spectra, we demonstrate that this phenomenon occurs when 
weak disorder is introduced in a RMT spectrum or when strong disorder acts in a Poisson regime.
We show that the long-range statistics follow the Taylor's law, which suggests the presence of a fluctuation 
scaling (FS) mechanism in this kind of spectra.
\end{abstract}

\maketitle

\section{Introduction} 

In the late 50s, E. Wigner proposed an ensemble of random matrices as a tool to describe statistical
properties in the dense region of the spectra of many-body systems. 
During the 60s, the formalism was then fully developed by Wigner himself, 
M. L. Mehta and mainly by Dyson in a series of seminal papers (see Ref.  \cite{Porter} for a review with preprints). 
Random Matrix Theory (RMT) 
could then be considered as a well established theory with a body of statistical measures that became known as the 
Wigner-Dyson statistics \cite{Meht}. This standard RMT is constituted by three classes of ensembles of Hermitian 
Gaussian matrices whose elements are real, for the Gaussian Orthogonal Ensemble (GOE), complex, for the Gaussian 
Unitary Ensemble (GUE) and quaternion, for the Gaussian Symplectic Ensemble (GSE). These classes are labelled by 
the Dyson index $\beta$ that gives the 
number of degree of freedom of the matrix elements, being 1, 2 and 4 respectively. A great boost 
in applications came in the beginning of the 80s when the link 
to the manifestations of chaos in quantum system was set by Bohigas-Giannoni-Scmit conjecture
that states the equivalence between quantum chaos and RMT \cite{Schmit} while, in contrast, regular systems
would have the uncorrelated Poisson statistics. The Wigner-Dyson statistics contains two kinds
of measures: short-range ones that probe local correlations, for which the most used measurement is the nearest-neighbor
distribution (NND) and the long-range ones that probe correlations along the spectrum, for which 
the number variance (NV) is the most employed quantity.   

Despite the success RMT had in more than half century of its existence (see the old review paper \cite{Guhr} with over 800
references), extensions of the original formalism have been proposed in order 
to enlarge the range of applications of the random matrix approach. 
Here we are particularly interested in three relatively recent RMT generalisations. The first one is the so-called 
$\beta$-ensemble \cite{Edelman} made of tridiagonal matrices,  
in which the value of the Dyson index $\beta$ can assume any real positive value in contrast to the values of 1, 2 and 4 it has in the Gaussian ensembles. The second generalisation 
has been called the disordered ensemble, in which an external source of randomness is introduced that operates concomitantly with the internal Gaussian ones \cite{disord}. Finally,
the third one is the ensemble constructed by randomly removing a fraction of
the eigenvalues from a given spectrum, namely the thinned ensemble \cite{Boh,Boh1}. The removal of levels decreases the
correlations among the remaining ones so that they show statistics intermediate between Wigner-Dyson and Poisson. 
It is our motivation to show that by combining these 
three RMT extensions, a random matrix model is constructed that can capture special features found in the analysis 
of spectra which are far from levels of physical systems. 

Spectra are points on a line, for instance the sequence of prime numbers form a spectrum. Moreover spectra can be also constituted in situations
in which line and points are considered in a general way \cite{Carpena}. For example, as studied in Ref. \cite{Deng},
 if punctuations are removed from a text, then it becomes a spectrum of blanks. In this case, the NND is
the distribution of the distances between  neighboring blanks, measured by the number of letters
and it gives the distribution of the length of words. The same idea can be applied to Chinese characters by considering that for characters strokes play the same role as letters do for words. For polymers, protein and DNA are sequences of letters, by just taking out a given letter, then a spectrum is defined. 

We have analysed the collection of data extracted from the above four areas of non-standard spectra by calculating 
both the local NND and the long-range NV. The need of the three RMT extensions is then immediately justified by the data. 
The three RMT classes, GOE, GUE and GSE, are associated with symmetries of the physical systems which play no role in 
the present case turning necessary to consider
arbitrary real values of $\beta$. We found that the NVs show a parabolic increase for large interval, a super-Poissonian behavior 
which is a
characteristic of disordered ensembles \cite{disord}. In some cases,the NND and the NV can only be fitted by resorting to the 
intermediate statistics of the thinned ensemble.
However, the most important feature emerges when the NND and NV data are confronted, it is found that they can 
behave independently.
This means that we are dealing with a special kind of spectra that shows a certain degree of complexity and elasticity.
 As these are typical characteristics of soft matter \cite{deGennes} and, also, we have datasets from soft material,
 we thought that it would be appropriate to use the expression soft spectra. 

The analogy with the state of matter is known in the theory of spectra \cite{Boh3}. Actually, the configuration
exhibited by the RMT eigenvalues as a consequence of the repulsion among them has already been 
considered as a crystal lattice structure. The picture is that the eigenvalues behave as a 
picket fence in which they vibrate around fixed points \cite{Boh2}. Here the analogy is extended to 
show that a new phase appears when spectra are subjected to an external source of randomness.

This paper is organized as follows: in the next section, we discuss in details of the three RMT extensions, the main points are 
illustrated on Figs. 1-3, in the Appendix A we show how the external source modifies the number variance and in Appendix B
we derive the basic asymptotic expressions for very long spectra. Section III applies the formalism in 
the analysis of spectra extracted in four different areas of quantitative linguistics and polymers (see Appendix C and D for details). 
These results are summarised in Tables I and II, and Figs. 4-7. We conclude the results in the last section.
 
\section{Disordered thinned beta ensemble} 
 
 Before combining the three RMT extensions, namely the $\beta$, the disordered and the thinned ensembles, we give a summary of their main points.
 
 \subsection{The $\beta$-ensemble}
 
 The $\beta$-ensemble consists in a family of Hermitian tridiagonal matrices 

\begin{equation}
H_{\beta} =\left(\begin{array}{c c c c c c}
a_1 & b_1  &          &          &         &\\
b_1 & a_2  & b_2      &          &         & \\
    &   .  & .        &          &         &\\
    &      &  .       &  .       &         & \\
    &      &          &  .       & .       & \\
    &      &          &  b_{N-2} & a_{N-1} & b_{N-1}  \\
    &      &          &          & b_{N-1} & a_{N} \\
\end{array}\right)  ,
\end{equation}
where the diagonal elements $a_i$  are normally distributed, namely
$N(0,1),$ while  the $b_i$'s are distributed according to

\begin{equation}
f_{\nu} (y)=\frac{2\exp(-y^2) y^{\nu-1}}{\Gamma(\nu/2)},
\end{equation}
with $\nu=(N-i)\beta$ and $\beta$ is a real positive parameter. From this definition, 
it is found that the joint density distribution of the eigenvalues is given by \cite{Edelman}

\begin{equation}
P_{\beta} (x_1 ,x_2 ,...,x_n)=C_{n} \exp\left(-\frac{1}{2}\sum_{k=1}^{n} x_{k}^2 \right)
\prod_{j>i}|x_j - x_i |^{\beta}.
\end{equation}
 As stated in the introduction, the above equation shows that for $\beta=1,2,$ and $4$ 
 their eigenvalues share all the statistical 
 properties of the RMT Gaussian classes of matrices, that is, they have Wigner-Dyson statistics. 
 For arbitrary values of $\beta,$ analytic expressions are not yet fully derived. 
 Asymptotically, it can be shown that when $\beta\rightarrow 0$ with the matrix size kept fixed,
 the matrix becomes diagonal, and in this case the density of eigenvalues is Gaussian 
 and the Poisson statistics follows. On the other hand, when $\beta\rightarrow\infty$ 
 fluctuations are suppressed.

 For the product $N\beta >> 1$, the asymptotic density of eigenvalues is the semi-circle law   
 
 \begin{equation}
\rho_{\beta}(\lambda)=\frac{1}{\pi\beta}\sqrt{2N\beta-\lambda^2},
\end{equation}
and the NND is well described by the Wigner surmise

\begin{equation}
p_{\beta}(s)=\frac{2 B^{\beta+1}s^{\beta}}{\Gamma[(\beta+1)/2]}\exp\left[-(Bs)^2\right],  \label{3}
\end{equation}
where $B=\Gamma(\frac{\beta+2.}{2})/\Gamma(\frac{\beta+1}{2})$ and $s=2N_{\beta}(\lambda/2)$ with

\begin{equation}
N_{\beta}(E)=\frac{N}{\pi}\left(\arcsin\frac{E}{\sqrt{2N\beta}}+\frac{E}{\sqrt{2N\beta}}\sqrt
{1-\frac{E^2}{2N\beta}}\right) . \label{Hbeta}
\end{equation}
Eq. (\ref{3}) defines an one parameter family of functions, whose parameter $\beta$ can be determined by
fitting the data as been done in Ref. \cite{Deng}. 

For $N\beta<<1,$ the density is 

\begin{equation}
\rho_{P} (\lambda)=\frac{N}{\sqrt{2\pi}} 
\exp\left(-\frac{\lambda^2}{2}\right), \label{7}
\end{equation}
which is that of the diagonal matrix elements, and in the ``unfolded" variable $s=2N_{P}(\lambda/2)$ with

\begin{equation}
N_{P}(\theta)=\frac{N}{\sqrt{2\pi}}\int_{0}^{\theta}dt\exp(-\frac{t^2}{2})=
\frac{N}{2}\mbox{erf}\left(\frac{\theta}{\sqrt{2}}\right), \label{NP1}
\end{equation}
the statistics is Poissonian, that is, the NND is $\exp(-s)$  and the NV is $\Sigma^2_{P} (L)=L$.
 
\subsection{Disordered ensemble} 

The disordered ensemble was defined in \cite{disord} by considering random matrices  
$H(\xi)$ which are obtained from a matrix $H$ of a given ensemble through the relation

\begin{equation}
H_{D}(\xi)=\frac{\bar\xi}{\xi}H,  \label{1} 
\end{equation}    
where $\xi$ is a positive random variable sorted from a distribution $w(\xi)$ with first moment ${\bar \xi}$. 
This scheme emerged from a generalization of an ensemble
generated, by two independent groups, using  the maximum entropy principle based on the  
Tsallis entropy \cite{Raul, Bertuola}. And it has been named disordered 
as an external source of randomness, which is represented by the random parameter $\xi$ imposed to the internal
randomness.   
The localization of the distribution $w(\xi)$ around its average controls the amplitude of the disorder.
 
If Eq. (\ref{1}) is diagonalized then it is found that a disordered eigenvalue $x_k$ of $H_D$ is  
related to an eigenvalue $E_k$ of $H$ as  $x_{k}=\sqrt{\frac{\bar\xi}{\xi}}E_k $ with $k=1,2,...N$. 
From this important relation, it follows  
that the joint density distribution of the disordered eigenvalues is expressed in terms of the joint
distribution of $H$ as

\begin{equation}
P_{D} (x_1,x_2,...,x_N;{\bar\xi})=\int_{0}^{\infty} d\xi w(\xi)\left(\frac{\xi}{\bar\xi}\right)^{N/2}
P\left(\sqrt{\frac{\xi}{\bar\xi}}x_1,
\sqrt{\frac{\xi }{\bar\xi}}x_2,...,\sqrt{\frac{\xi }{\bar\xi}}x_N\right).
\end{equation}
Therefore, in order to construct the disordered $\beta$-ensemble, we generalize the original scheme  by making,
in Eq. (\ref{1}), the identification  of $H$ with the $H_{\beta}$.  

\subsubsection{Effect of disorder in the RMT regime $N\beta>>1$}

For the case $N\beta >>1$, the one-point function, that is the density, is obtained 
by integrating all eigenvalues and multiplying by $N,$ this gives

\begin{equation}
\rho_{D\beta} ( x) =\int_{0}^{\infty}d\xi w(\xi)\left(\frac{\xi}{\bar{\xi}}\right)^{1/2}
\rho_{\beta}\left(\sqrt{\frac{\xi}{\bar\xi}}x\right)=
\frac{1} {\pi\beta}\int_{0}^{\xi_{max}}
d\xi w( \xi) \left(\frac{\xi}{\bar{\xi}}\right)^{1/2}\sqrt{2N\beta -
\frac{\xi}{\bar{\xi}}x^2 }, \label{126}
\end{equation}
where $\xi_{max}=2N\beta\bar{\xi}/x^2$. 
Integrating the density, Eq. (\ref{126}), from the origin to a value $x$,
the cumulative function is obtained as

\begin{equation}
N_{D\beta} (x)=\frac{N}{2}\left(1-\int_0^{\xi_{max}}d\xi w(\xi)\left[1-\frac{2}{N}
N_{\beta}\left(\sqrt{\frac{\xi}{\bar{\xi}}}x \right)\right]\right) \label{141}. \label{D1}
\end{equation}

To measure the short-range spectral fluctuations we define
the probability $E(s)$ that the interval $(-\frac{s}{2},\frac{s}{2})$
is empty. This so-called gap probability is obtained by integrating
over all eigenvalues outside the interval 
$(-\frac{\theta}{2},\frac{\theta}{2})$ to obtain

\begin{equation}
E_{D\beta}(s)=\int_0^{\infty}d\xi w(\xi) E_{\beta}\left(2N_{\beta}\left(
\sqrt{\frac{\xi}{\bar{\xi}}}\frac{\theta}{2}\right)\right], \label{108s}
\end{equation}
with $s=2N_{D\beta}(\frac{\theta}{2})$.
From the gap probability, the NND is obtained by taking the derivatives, $F(s)=\frac{dE}{ds}$ 
and $p(s)=\frac{d^{2}E}{ds^2},$ such that the $p(s)$ is the distribution and $1+F(s)$ the 
probability. Analytic expressions 
for $E_{\beta}(s)$ are only known for the Gaussian ensemble, in particular,
for the case of $\beta=1$, we use the Wigner surmise 
\begin{equation}
E_1 (s)=\mbox{erfc} (\sqrt{\frac{\pi }{4}}s )  .
\end{equation}
The number variance of the disordered ensemble is given by (see the derivation in the Appendix A)

\begin{equation} 
\Sigma^2_{D\beta} (L) = <n^2> - <n>^2 =\int_{0}^{\infty} d\xi w
\left(\xi \right) \left[
\Sigma^2_{\beta} \left(2N_{\beta}(\sqrt{\frac{\xi}{\bar{\xi}}}\frac{\theta}{2})\right)  + 4N_{\beta} ^2
\left(\sqrt{\frac{\xi}{\bar{\xi}}}\frac{\theta}{2} \right)  \right] - L^2 ,\label{37}
\end{equation}
where $L=2N_{D\beta}(\frac{\theta}{2}).$

To proceed and be able to analyze the situation of weak disorder, we start by choosing 
the distribution $w(\xi)$ to be given by (see \cite{Abul,Hmodel} for other choices)

\begin{equation}
w(\xi)=\frac{1}{\Gamma({\bar \xi})}\exp(-\xi)\xi^{{\bar \xi}-1} \label{127}, \label{Ts}
\end{equation}
which follows the Tsallis entropy formalism. Using the results of Appendix B, we find 
that asymptotically the number variance takes the simple form of a parabola given by

\begin{equation} 
\Sigma^2_{D\beta} (L)\simeq  \left[\left(\frac{\overline{\sqrt{\xi}}}
{\sqrt{\overline{\xi}}}\right)^{-2} -1\right]L^2 \simeq  \frac{1}{4\bar\xi}L^2, \label{37}
\end{equation}
where the $\Sigma^2_{\beta}$ term that increases logarithmically, 
in the presence of the $L^2$ term, has been neglected. Therefore, in this case the number variance
satisfies the Taylor's law \cite{Taylor} exhibiting the phenomenon that is more recently named as the fluctuation scaling \cite{Kertesz,Eisler,Menezes}.

On the other hand, considering the gap probability, the first order term in Eq. (\ref{B5}) 
can be used such that with $s\simeq \rho(0)\theta$ being replaced in Eq. (\ref{108s}), it becomes

\begin{equation}
E_{D\beta}(s)=\int_0^{\infty}d\xi w(\xi) E_{\beta}\left(
\sqrt{\frac{\xi}{\bar{\xi}}}s\right)\simeq  E_{\beta}(s), 
\end{equation}
as the disorder fluctuations are quenched in the large $\bar\xi$ limit.

The above results show that, asymptotically, the two sources of randomness
acting on the system affect, differently, the short and the long range statistics. The local statistics
are described by the expressions of the $\beta$-ensemble, while the long-range number variance is dominated
by the external source of randomness. This important result is illustrated by the numerical simulations exhibited
in Fig. 1, it clearly shows the robustness of the 
local statistics in contrast with the high sensitivity of the long range one.

\subsubsection{Effect of disorder in the Poisson limit $N\beta<<1$}

Considering now the Poisson limit of the $\beta$-ensemble, for the disordered density the
expression

\begin{equation}
\rho_{DP} (\lambda)=\frac{N}{\sqrt{2\pi}}\int_{0}^{\infty}w(\xi)\sqrt{\frac{\xi}{\bar\xi}}
\exp\left(-\frac{\xi\lambda^2}{2\bar\xi}\right)=\frac{N\Gamma(\bar\xi +1/2)}
{\sqrt{2\pi\bar\xi}\Gamma(\bar\xi)}\left(1+\frac{\lambda^2}{2\bar\xi}\right)^{-\bar\xi-1/2}
\end{equation}
is derived where Eq. (\ref{7}) was used. From this equation, the cumulative function  

\begin{equation}
N_{DP}(\lambda)=\int_{0}^{\lambda}dt\rho_{DP} (t)=\frac{N\Gamma(\bar\xi +1/2)\sqrt{\bar\xi}}
{2\sqrt{\pi\bar\xi}\Gamma(\bar\xi)}
B\left(\frac{\lambda^2}{2\bar\xi};\frac{1}{2},\bar\xi + \frac{1}{2}\right) 
\end{equation}
is obtained, where $B(x;a,b)$ is the incomplete beta function. For the gap probability, 
explicitly we have

\begin{equation}
E_{DP}(s)=\int_0^{\infty}d\xi w(\xi) E_{P}\left[2N_{P}\left(
\sqrt{\frac{\xi}{\bar{\xi}}}\frac{\theta}{2}\right)\right] = 
\int_0^{\infty}d\xi w(\xi) \exp\left[-2N_{P}\left(
\sqrt{\frac{\xi}{\bar{\xi}}}\frac{\theta}{2}\right)\right] , \label{117}
\end{equation}
where $s=2N_{DP}(\theta/2).$

Using the straight line expression for the Poisson number variance, the disordered NV is then given by

\begin{equation} 
\Sigma_{DP}^2 (L) =\int_{0}^{\infty} d\xi w
\left(\xi \right) \left[
2N_P(\sqrt{\frac{\xi}{\bar{\xi}}}\frac{\theta}{2})  + 4N_P ^2
\left(\sqrt{\frac{\xi}{\bar{\xi}}}\frac{\theta}{2} \right)  \right] - L^2 =
L+4\int_{0}^{\infty} d\xi w
\left(\xi \right) N_P ^2
\left(\sqrt{\frac{\xi}{\bar{\xi}}}\frac{\theta}{2} \right)  -L^2  ,\label{41}
\end{equation}
where $L=2N_{DP}(\frac{\theta}{2})$. 

The above equations are exact, considering now that we are dealing with very long spectra, we can make
the same analysis of the asymptotic behavior of the expressions. Here we would like to make $N$ goes to infinity
in Eq. (\ref{NP1}) keeping the product $N\theta$ finite, immediately, we derive that, as before, 
the cumulative function can be approximated as $N_{P}\sim\rho_{P}(0)\theta$. Using then results of Appendix B, 
we find that the number variance satisfies the more general expression of Taylor's law \cite{Eisler}

\begin{equation} 
\Sigma^2_{DP} (L)\simeq  L + 
\left[\left(\frac{\overline{\sqrt{\xi}}}{\sqrt{\overline{\xi}}}\right)^{-2} -1\right]L^2 , \label{P13}
\end{equation}
in which the linear Poisson term competes with the superPoisson term.

Turning to the gap probability, introducing 
$s\simeq  \rho_{P}(0) \frac{\overline{\sqrt{\xi}}}{\sqrt{\overline{\xi}}}\theta$
in Eq. (\ref{117}), it can be rewritten as

\begin{equation}
E_{DP}(s)\simeq 
\int_0^{\infty}d\xi w(\xi) \exp\left[-
\frac{\sqrt{\xi}}{\overline{\sqrt{\xi}}} s\right] ,
\end{equation}
and the NND

\begin{equation}
p_{DP}(s)\simeq \int_0^{\infty}d\xi w(\xi) \left(\frac{\sqrt{\xi}}{\overline{\sqrt{\xi}}}\right)^{2}
 \exp\left[-\frac{\sqrt{\xi}}{\overline{\sqrt{\xi}}} s\right] =\frac{k^{2}
 \Gamma(2\bar\xi +2)}{2^{\bar\xi}\Gamma(\bar\xi)}
\exp\left(\frac{k^{2}s^{2}}{8}\right)
U\left(2\bar\xi +\frac{3}{2},\frac{ks}{\sqrt{2}}\right), \label{P11}
\end{equation}

where $k=(\overline{\sqrt{\xi}})^{-1}$ and $U(a,x)$ is the parabolic cylinder function\cite{Abram}.
With this identification, immediately we can resort to the asymptotic of $U(a,x),$ thus if
$a>>x$, that is, if $\bar\xi >>s$ we find that $p_{DP}\simeq  p_{P}=exp(-s)$ which is the Poisson
limit. On the other hand, for moderate values of $a,$ the Dawson expansion in $\sqrt{a^2 + x^2}$
shows deviations of the Poisson regime with an appearance of a power law decay. 

In Fig. 2, it is shown the effect of disorder in the short and long-range statistics in the Poisson case. 
We observe that, for relatively small value $\bar\xi =2$ the gap probability is still close to Poisson 
distribution, but it starts to show deviations. In contrast, the number variance shows large deviations 
from the linear Poison behavior.

It is important to observe that when the value of the parameter $\bar\xi$ approaches to one, a strong disorder regime is reached in which the local statistics show a 
power-law decay \cite{Bertuola,disord}.

\begin{figure}[ht]
 \centerline{\includegraphics*[angle=0,width=0.50\textwidth]{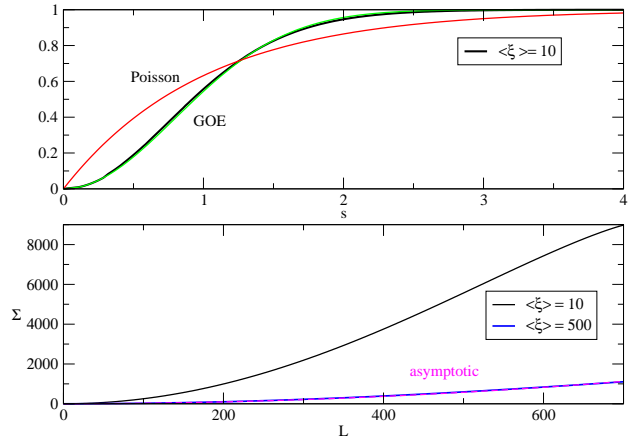}}
 \caption{The effect of disorder on the GOE ($\beta=1$) cumulative nearest-neighbor 
 distribution and on the number variance. It shows the robustness of the local statistics in 
 contrast to the sensitivity of the long range one.}
\end{figure}

\begin{figure}[ht]
 \centerline{\includegraphics*[angle=0,width=0.50\textwidth]{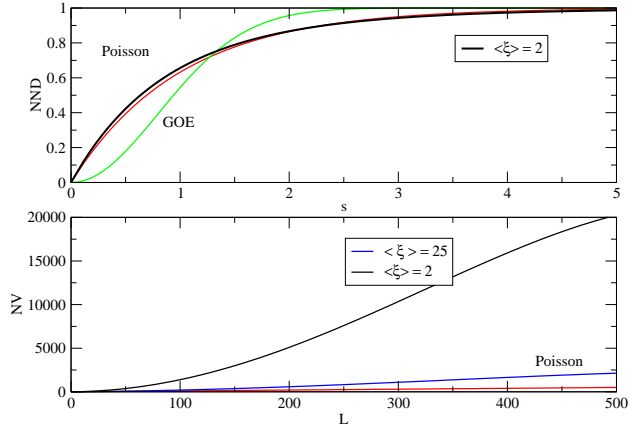}}
 \caption{The effect of disorder on the Poisson cumulative nearest-neighbor distribution
 and on its number variance. One can observe that when the parameter $\bar\xi$ approaches to one, 
 a strong disorder regime is reached in which the local statistics show a power-law decay.}
\end{figure}

\subsection{Thinned ensemble}

Using the $\beta$-ensemble, we are supposing that we are dealing with perfect spectra which, 
necessarily, is not the case for the ones we are interested in. In order to make our approach more
comprehensive, we add to the model the formalism introduced in \cite{Boh1} that deals with 
incompleteness in a sequence being analyzed. This formalism was a natural development of 
the missing level theory \cite{Boh} and it consists in constructing from a given spectrum a new one
by removing with a probability $1-f$ levels from it, such that the resulting spectrum has, 
in average, $f$ levels. In statistics, this construction is denoted as a thinning point 
process \cite{thinning} and it has 
the important aspect of preserving, in the less dense object, properties of the original one. 
The RMT formalism is based on Fredholm determinants \cite{Meht} and this determinantal method
is preserved by the thinning process. This fact, 
explain the great attraction recently arose by this model \cite{Chen, Deift, Duits, Texier,
Bothner}  

In \cite{Boh1}, it is shown that the thinned spectra has statistics 
intermediate between RMT and Poisson. Moreover, the RMT formalism analytically also 
describes this intermediate situation with $f$ playing the role of 
a parameter that varies from zero to one. 
In terms of the spacing distribution of the initial spectrum the NND is given by

\begin{equation}
p(s,f)=\sum_{k=0}^{\infty }\left( 1-f\right) ^{k}P(k,\frac{s}{f}) ,
\end{equation}
where small case denotes the thinned quantity and capital the original one. The $P(k,s)$ are spacing
distribution with $k$ levels inside the interval $s$ and the division by $f$ taking into account the contraction
of the incomplete spectrum. Since in (25), all spacing distributions $P(k,s)$ are normalized with first moments 
$< s >_{k}= k + 1,$ it then follows that $p(s, f)$ also is normalized and has first moment equal to one. 
This shows the need of rescaling the argument of the functions with the parameter $f$, accordingly for the density we have 

\begin{equation}
\rho_{\beta}(\lambda,f)=f\rho_{\beta}(\lambda),
\end{equation}
while for the number variance it can be shown that \cite{Boh1}

\begin{equation}
\sigma_{\beta} ^{2}\left( L,f\right) =\left( 1-f\right) L+f^{2}\Sigma_{\beta} ^{2}
\left( \frac{L}{f}\right) . \label{64}
\end{equation}
The NND expression implies that the thinned spectra still have level repulsion and
an exponential decay as can be seen in the Fig. 3. On the other hand, the number variance expression shows 
an asymptotically straight line Poisson behavior. Finally, we remark that the thinning process has no effect
on a uncorrelated Poisson spectrum.  

\subsection{The combined ensemble}

Combining these three ingredients, we have a disordered thinned $\beta$-spectrum. Numerical simulations show that
the local statistics are very robust with respect to disorder while the long range statistics, in contrast,
are very sensitive. As a consequence, we have found that the NND data of the spectra of blanks can be fitted
considering only the thinning effect in the $\beta$-ensemble spectra. 
But for the number of variance data better results are obtained using
the expression that takes into account of effects of disorder and thinning that, explicitly, leads to 

\begin{equation} 
\Sigma^2_{D\beta T} (L)  =\int_{0}^{\infty} d\xi w
\left(\xi \right) \left[
\sigma^2 \left(2N_{\beta}(\sqrt{\frac{\xi}{\bar{\xi}}}\frac{\theta}{2}),f\right)  + \frac{4}{f^2}N_{\beta} ^2
\left(\sqrt{\frac{\xi}{\bar{\xi}}}\frac{\theta}{2} \right)  \right] - \left(\frac{L}{f}\right)^2 ,\label{41}
\end{equation}
where $L=2N_{D\beta}(\frac{\theta}{2})$ such that, asymptotically, we have

\begin{equation} 
\Sigma^2_{D\beta T} (L)\simeq (1-f)L +  \left[\left(\frac{\overline{\sqrt{\xi}}}
{\sqrt{\overline{\xi}}}\right)^{-2} -1\right]\left(\frac{L}{f}\right)^2. \label{37q}
\end{equation}

In Fig. 3, the effect of thinning is shown: the NND shows a typical intermediate 
statistics behavior while an increase in the curvature of the super-Poissonian parabola occurs in the NV.  

\begin{figure}[ht]
 \centerline{\includegraphics*[angle=0,width=0.50\textwidth]{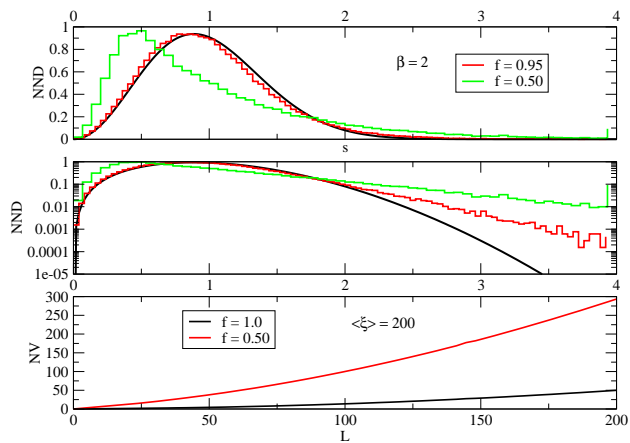}}
 \caption{The effect of removing levels on the GUE ($\beta=2$) cumulative nearest-neighbor distribution and on the number variance. 
 The nearest-neighbor distribution implies that the thinned spectra still have level repulsion and an exponential decay, while the 
 number variance, asymptotically, shows a typical straight line behavior.}
\end{figure}

Therefore, in all cases we have NVs that follow the Taylor's law in the form $aL + bL^2,$ this implies in the presence
of a FS phenomenon. As a matter of fact, here the fluctuation in the scaling can be understood as 
the manifestation of the breaking of the ergodicity of the ensemble enacted by the introduction of the external 
randomness \cite{disord}. Ergodicity in RMT means that averaging over one large matrix is equivalent to an ensemble average,
in other words, individual matrices are equivalent which, by construction, is not the case of the disorder ensemble.

%%\begin{figure}[ht]
%% \centerline{\includegraphics*[angle=0,width=0.50\textwidth]{sig1.eps}}
%% \caption{The combined effect of disorder and removing levels.}
%%\end{figure}

\section{Applications}

We have applied the formalism in the analysis of spectra extracted from four different areas, namely,
spectra of blanks in Portuguese literary texts, spectra of space between strokes of Chinese characters 
in Chinese literature, spectra of protein and spectra from DNA sequences, please find the description of the datasets in Appendix C. In the main text, we only show the results of four samples for each system, results of more samples are shown in Appendix D. Tables I and II show that the average length $\langle L \rangle$ 
of the ``words" is a characteristic 
quantity for each sequence. Although the results of Protein and DNA have similar behavior, the $\langle L \rangle$ are 
different from each other. One point we would like to stress is that the length of the sequence being analyzed should be 
long enough otherwise the data could lead to misleading results. Another important point is that we are considering spectra 
for which an average constant density can be assumed. This allows to rescale the data in order to have the so-called unfolding, 
namely, an average spacing equals to one.

\subsection{Fitting procedure}

For Wigner-like family, we have the specific equations (\ref{3}) and (\ref{37q}) to fit, respectively, the NNDs and NVs by using the least square fitting method. The values of the fitting parameters and fitting quality parameters are shown in Table I. 

For Poisson-like family, the specific fitting formulas to be used are (\ref{P11}) for NNDs and (\ref{P13}) for NVs, 
the values obtained of the fitting parameters and fitting quality parameters are shown in Table II.

\begin{table*}%[h!]
\begin{center}
\caption{Values of the basic parameters for each sequence of the Wigner-like family. $\langle L \rangle$ is the average length, $\beta$ is the parameter of $\beta$-ensemble.  $f_{NV}$ is the fitting parameters of number variance for the thinned ensemble. $\langle \xi_{NV} \rangle$ are the fitting parameters of NND and number variance for the averages of $\xi$ sorted from a distribution $w(\xi)$ in the disordered ensemble. Values of $R_{NND}^{2}$ and $R_{NV}^{2}$ show the fitting qualities of NND and number variance. }

\begin{tabular}{p{2.5cm}p{2.5cm}p{1.5cm}p{1.5cm}p{1.5cm}p{1.5cm}p{1.5cm}p{1.5cm}}
\hline\hline
Area     &  Sequence & $\langle L \rangle$ & $\beta$ &  $f_{NV}$ & $\langle \xi_{NV} \rangle$ &  $R_{NND}^{2}$ & $R_{NV}^{2}$ \\
\hline
\multirow{3}{*}{Portuguese} & AFilha & 4.615 & 0.486  &0.791 &323.246  & 0.969 &0.9998  \\
 & Grande &4.501 & 0.652  &0.852  &  1720.11  &0.963 &0.9998  \\
& OPrimo & 4.828 &0.476  &0.802  &  1120.68   &0.975 &0.9998   \\
& OsMaias & 4.761 &0.44 &0.839   &  1486.1  &0.983 &0.9999 \\
\hline
\multirow{3}{*}{Chinese} & HLM & 7.019 &1.433 &0.821  &474.174 & 0.981 &0.9999 \\
& SHZ & 7.169 &1.627 &0.835 & 666.515 &0.977 &0.99998 \\
& PFD & 7.021 &1.792 &0.843 & 927.198 &0.988 &0.99997 \\
& SHL & 6.937 &1.869 &0.860 & 2566.82 &0.972 & 0.9999 \\
\hline\hline
\label{tab1}
\end{tabular}
\end{center}
\end{table*}

\begin{table*}%[h!]
\begin{center}
\caption{Values of the basic parameters for each sequence of the Poisson-like family. $\langle L \rangle$ is the average length. $\langle \xi_{NND} \rangle$ and $\langle \xi_{NV} \rangle$ are the fitting parameters of NND and number variance for the averages of $\xi$ sorted from a distribution $w(\xi)$ in the disordered ensemble. Values of $R_{NND}^{2}$ and $R_{NV}^{2}$ show the fitting qualities of NND and number variance. }

\begin{tabular}{p{3.0cm}p{3.0cm}p{1.5cm}p{1.5cm}p{1.5cm}p{1.5cm}p{1.5cm}}
\hline\hline
Area     &  Sequence & $\langle L \rangle$  & $\langle \xi_{NND} \rangle$ & $\langle \xi_{NV} \rangle$ &  $R_{NND}^{2}$ & $R_{NV}^{2}$ \\
\hline
\multirow{3}{*}{Protein} & A2ASS6 & 9.764  &1.274 &2.441 &0.946 &0.998 \\
& G4SLH0 & 6.823  &1.303 &4.099 &0.776 &0.995 \\
& Q8WZ42 & 9.468 &1.274 &2.415 &0.956 &0.9997 \\
& Q9I7U4 & 5.663 &1.233 &3.365 &0.901 &0.996 \\
\hline
\multirow{3}{*}{DNA} & A2ASS6 & 17.063 &1.277  &2.45 &0.953 &0.997 \\
& G4SLH0 & 11.039  &1.237  &2.12 &0.759 &0.988 \\
& Q8WZ42 & 13.937  &1.272  &2.03 &0.945 &0.9996 \\
& Q9I7U4 & 11.139 &1.224  &1.328 &0.839 &0.997 \\
\hline\hline
\label{tab1}
\end{tabular}
\end{center}
\end{table*}

\subsection{Spectra of blanks}

In \cite{Deng}, the spectra of blanks of literary texts of ten languages were analyzed and two
families have been found. The family denoted Poisson-like were fitted with a displaced Poisson distribution
as short words did not show a clear statistical behavior. For this reason, we are not considering here spectra from 
this family and decided to perform a re-analysis of the spectra of Portuguese, a language of the Wigner-like family.  The four 
literary texts are the same of Ref. \cite{Deng} and the results for NNDs and number variances are presented 
in Fig. 4, respectively.

We start by fitting the NND of each text with results shown in the Table.  Although, they are fitted with different values of the
parameter $\beta$ the data show that the set of NNDs can be considered as just fluctuations around an average distribution.
This remarkable feature suggests that the local statistics, that is the NND, does not
distinguish the spectra of the four texts. In contrast,  while the NV data show a clear separation.  They show a very good
individual fit in agreement with the theoretically predicted parabolic Taylor's law.

\begin{figure}[ht]
 \includegraphics*[angle=0,width=0.50\textwidth]{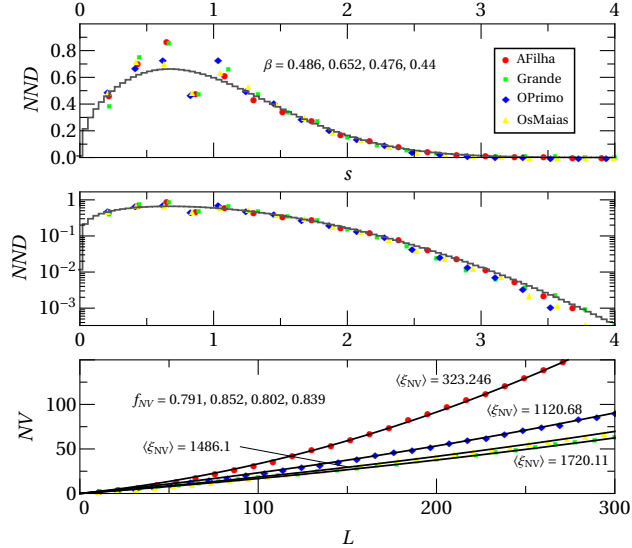}
 \caption{The NND and number variance of Portuguese. For NND, each one can be fitted using only the $\beta$-ensemble with 
values of the $\beta$ shown, the solid line correspond to the average value  $\beta=0.486$ and it shows that the four NNDs 
fluctuate around an average NND.This is confirmed in log-scale plot. Each number variance shows a super-Poisson behavior and
a perfect agreement with a parabolic Taylor's law.}
\end{figure}

\subsection{Chinese characters}

The writing system is the process or result of recording spoken language using a system of visual marks on a surface. There are mainly two types of writing systems: logographic (Sumerian cuneiforms, Egyptian hieroglyphs, Chinese characters) and phonographic. The latter includes syllabic writing (e.g. Japanese hiragana) and alphabetic writing (English, Russian, Portuguese), while the former encodes syllables and phonemes.

The unit of Chinese writing system is the character: a spatially marked pattern of strokes. In ideogram the stroke plays the similar role as the letter does in the alphabetic language. We investigated four Chinese texts: \begin{CJK*}{UTF8}{gbsn}
三国演义, 水浒传, 平凡的世界, 撒哈拉沙漠的故事, which are denoted as HLM, SHZ, PFD and SHL. For example, the character ``生" has 5 strokes, namely `` 丿, 一, 一, 丨, 一".
\end{CJK*}

As it happened in the Portuguese case,  NND data fluctuate around an average NND suggesting it s a characteristic of the Chinese writing system common to the four texts. On the other hand, the NVs are well separated and can be fitted using a parabolic Taylor's law.  The NVs two first data, namely HLM and SHZ, were extracted from ancient classical novels while the two others come from
modern novels. Their NVs suggest that the modern ones tend to be more correlated that the old ones. 

\begin{figure}[ht]
\includegraphics*[angle=0,width=0.50\textwidth]{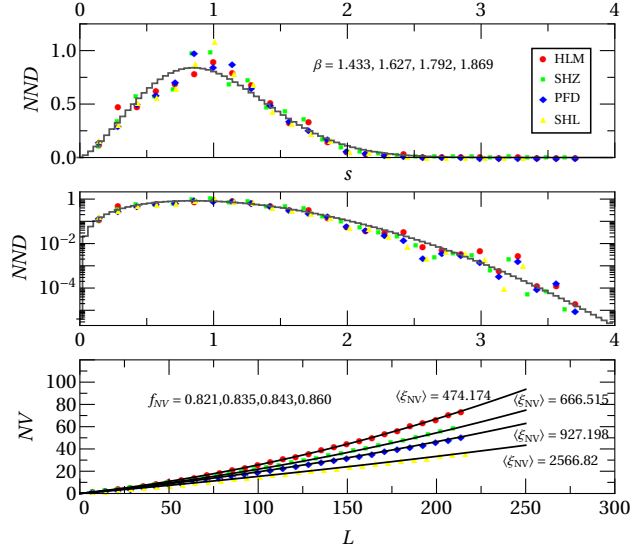}
 \caption{The NND and number variance of Chinese characters. As in the Portuguese case, each NND can be fitted with a particular value of $\beta$ but
thei data can be considered as fluctuations around an average distribution. For number variance a parabolic Taylor's law give very good fit, The 
data suggest that the modern ones (PFD and SHL) tend to be more correlated than the old ones (HLM and SHZ).}
\end{figure}

\subsection{Protein}

Protein has been analysed as the sequence within the statistical physics scope over the past few years, in which each symbol of the alphabet corresponds to one amino acid. There are 20 different amino acids in the protein sequence. We randomly used four proteins in the Swiss-Prot protein database for our analysis, which the primary accession number of these proteins are A2ASS6, Q8WZ42, G4SLH0 and Q9I7U4 respectively. We denote the most frequent amino acid of a protein as the blank, then get the NNDs and the number variances. 

It is shown that the NNDs simply can be fitted with the disordered Poisson distribution, while the number variances show super-Poisson behavior. 
These results indicate that this kind of spectral analysis works quite well in case of proteins, and therefore it would be a new method for protein analysis. For example, giving a amino acid in a protein sequence, we would know how the amino acid is distributed along the sequence.

\begin{figure}[ht]
 \centerline{\includegraphics*[angle=0,width=0.50\textwidth]{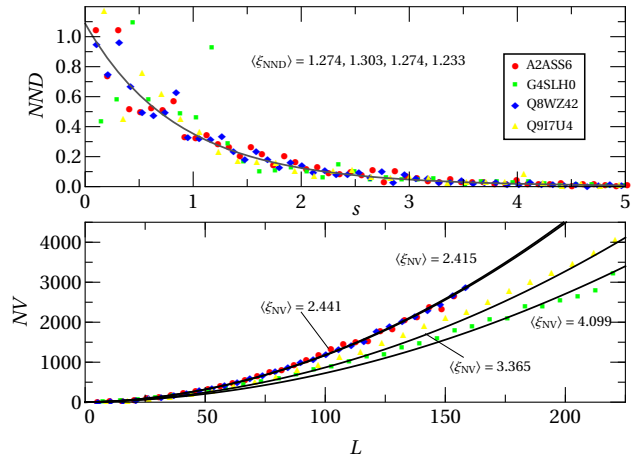}}
 \caption{The NND and number variance of protein. Each NND can be fitted using Eq. (\ref{P11}) but the figure also shows that they can be considered as fluctuations around an average distribution which is greater than one at the origin as has a power-law decay. Each number variance shows an
 agreement with a parabolic Taylor's law.}
\end{figure}

\subsection{DNA}

DNA contains all the information that organisms need to live and
reproduce themselves including the protein synthesis. There are four nucleotides (A,T,C,G) and thus we can consider the alphabets are composed by four symbols in DNA sequences. Each DNA sequence can only get a unique protein, but each protein can be coded by several different DNA sequences. To compare the results of proteins, we used one of the DNA sequence corresponding to a protein mentioned above, in which the primary accession numbers are A2ASS6, Q8WZ42, G4SLHO and Q9I7U4 respectively. We selected the most frequent letter of a DNA sequence as a spectrum of blanks to be analysed. For such sequences we get the NNDs and number variances.

The NNDs were fitted using the same disordered Poisson model of the protein case. The average NND
goes beyond Poisson at the origin and has a power-law decay for large separations. DNA is converted to messenger RNA (mRNA) during transcription and then during translation, mRNA is converted to protein, therefore it is natural that the DNA and protein would have similar results.

\begin{figure}[ht]
 \centerline{\includegraphics*[angle=0,width=0.50\textwidth]{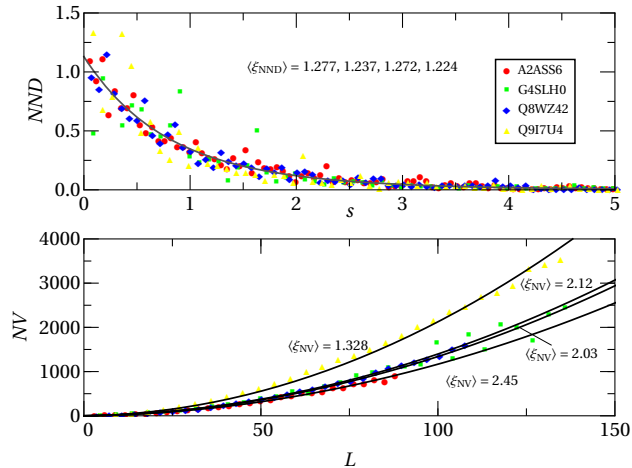}}
 \caption{The NND and number variance of DNA. They have the similar behavior of Protein. }
\end{figure}

\section{Conclusion}

We have analyzed spectra from four different areas that, in contrast with which is usual, do not belong to the Hamiltonian physical systems. From the results two main aspects are stuck out: first, the relative independence 
between the local ($NND$) and long-range ($NV$) statistics, and second, $NV$s of the parabolic form $aL + bL^2$ which is a characteristic of a fluctuation scaling, that is the FS-phenomenon. These two kinds of behaviors are not independent. In fact, considering that linear NV is a typical behavior of 
an uncorrelated spectrum (Poisson), we can infer that for relative short-range, the points of the spectra appear  
as an independent sequence, although the distances between neighbors follow a given NND (see note \cite{Poisson}).
However, as the range of observation increases, there is a crossover from the linear to the square dependence of the mean in the NV. This can be understood as been caused by the presence of an external driving that creates the long-range inhomogeneity responsible for the FS.

The analysis were done by fitting the results which were obtained by averaging along the empirical spectra (`time' average) with the analytic expressions provided by performing in the RMT model an ensemble average.     
The model combines three generalizations of the classical RMT ensembles: 
the beta ensemble, the disordered ensemble and the thinned ensemble, in
which statistic measures lie intermediate between Poisson and RMT. The disordered and thinned ensembles were developed as the extensions of standard RMT, but here we show that they also work in the beta case. In particular, in the limit when $\beta\rightarrow 0$, it gives origin to a disordered Poisson model 
that is proved to be a useful tool to describe the protein and DNA data. For the data associated to languages, disorder and 
thinning were essential to fit the data while in the case of protein and DNA thinning plays no role.
From the results obtained we conclude that the local statistics, that is the NND, is a characteristic of one area while 
the long-range statistics, that is the NV, can make distinctions of cases inside the area. In the linguistics cases,
the distinctions of NVs suggest that they can be associated to different styles and genres. From the NND results, we learn that language can
show level repulsion such that `levels' cannot cluster, however, the levels of Protein and DNA can cluster. Another
observation is that Protein and DNA can show power-law decay in their NND, which has been recently reported in the frequency of symbols \cite{Martin}.  

Finally we remark that we have modeled the disorder using the one parameter distribution, Eq.(\ref{127}). 
It is important to mention that other distributions have already been proposed. In Ref. \cite{Vivo}, for instance, another one parameter distribution was proposed and, a more general one, with three parameters 
was discussed in Ref. \cite{Hmodel}. It would be interesting to investigate if these families can, in some cases, provide more efficient fittings.

\section{Acknowledgments}

M.P.P. was supported by grant 307807/2017-7 of the Brazilian agency Conselho Nacional de Desenvolvimento Científico e Tecnológico (CNPq) and
is a member of the Brazilian National Institute of Science and Technology-Quantum Information (INCT-IQ). This work was supported in part by National Natural Science Foundation of China (Grant No. 11505071, 11747135, 11905163), the Programme of Introducing Talents of Discipline to Universities under Grant No. B08033, the Fundamental Research Funds for the Central Universities.

\appendix

\section{Number variance} 

To calculate the number variance we start with the expression \cite{Meht}

\begin{equation}
< n^{2} >_{G} = \int_{-\theta/2}^{\theta/2}dE_1\int_{-\theta/2}^{\theta/2}dE_{2} R_{2}(E_1,E_2)+
\int_{-\theta/2}^{\theta/2}dE \rho_{G}(E) ,
\end{equation}
for the average of the square! of the number of eigenvalues in the interval 
$(-\frac{\theta}{2},\frac{\theta}{2})$, where $R_{2}(E_1,E_2)$ is the two-point function.
Introducing disorder this quantity becomes 

\begin{equation}
< n^{2} > =\int_{0}^{\infty}d\xi w(\xi)\left[ \int_{-\theta/2}^{\theta/2}\int_{-\theta/2}^{\theta/2}
\frac{dx_{1}dx_{2}\xi}{\bar\xi} R_{2}\left(\sqrt{\frac{\xi}{\bar\xi}}x_1,\sqrt{\frac{\xi}{\bar\xi}}x_2\right)+
\int_{-\theta/2}^{\theta/2}dx\sqrt{\frac{\xi }{\bar\xi}} \rho_{G}\left(\sqrt{\frac{\xi }{\bar\xi}}x\right)\right] .
\end{equation}
Making in the integrals the substitution of variable $E=\sqrt{\frac{\xi }{\bar\xi}}x$, the above expression becomes

\begin{equation}
< n^{2} > =\int_{0}^{\infty}d\xi w(\xi)\left[ \int_{-\sqrt{\frac{\xi }{\bar\xi}}\theta/2}^{\sqrt{\frac{\xi }{\bar\xi}}
\theta/2}\int_{-\sqrt{\frac{\xi }{\bar\xi}}\theta/2}^{\sqrt{\frac{\xi }{\bar\xi}}\theta/2}
dE_{1}dE_{2} R_{2}\left(E_1,E_2\right)+
\int_{-\sqrt{\frac{\xi }{\bar\xi}}\theta/2}^{\sqrt{\frac{\xi }{\bar\xi}}\theta/2}
dE \rho_{G}\left(E\right)\right] .
\end{equation}
Changing variable as $t(E)=\int_{0}^{E}de^{\prime}\rho_{G}(E^{\prime})=N_{G}(E)$ and using that

\begin{equation}
\frac{R(E_1,E_2)}{\rho_{G}(E_1)\rho_{G}(E_2)}=1-Y(|t_2 - t_1|),
\end{equation}
we obtain

\begin{equation}
< n^{2} > =\int_{0}^{\infty}d\xi w(\xi)\left[-2\int_{0}^{N(\sqrt{\frac{\xi }{\bar\xi}}\theta/2)}
dt\left(1- 2N(\sqrt{\frac{\xi }{\bar\xi}}\theta/2))\right)Y(t) 
+ 2N_{G}\left(\sqrt{\frac{\xi }{\bar\xi}}\theta/2\right)
+4N_{G}^2\left(\sqrt{\frac{\xi}{\bar{\xi}}}\frac{\theta}{2} \right) \right] ,
\end{equation}
where the two first terms inside the brackets are just the number variance expression of the Gaussian
ensemble.

\section{Asymptotic expressions}

We are interested in the case of long spectra containing a large number of points. In this situation 
of very large $N,$ the statistics are measured around the center of the spectra. To be specific,
we want to make $N$ to go to infinity, in Eq. {\ref{Hbeta}), keeping the product $\sqrt{N}E$ finite, 
explicitly

\begin{equation}
N_{\beta}(E)=\frac{N}{\pi}\left(\arcsin\frac{\sqrt{N}E}{N\sqrt{2\beta}}+\frac{\sqrt{N}E}{N
\sqrt{2\beta}}\sqrt{1-\frac{NE^2}{2N^2\beta}}\right)\simeq \rho_{\beta}(0)E. \label{BB1}
\end{equation}
Assuming now that in Eq. (\ref{D1})  $\xi_{max}$ is very large and can be replaced by infinity, the disordered
cumulative function becomes 

\begin{equation}
N_{D\beta}(x)=\int_{0}^{\infty}d\xi w(\xi)N_{\beta}\left(\sqrt{\frac{\xi}{\bar{\xi}}}x\right)\simeq
\rho_{\beta}(0)\frac{\overline{\sqrt{\xi}}}{\sqrt{\overline{\xi}}}x,
\end{equation}
where (\ref{BB1}) has been used. Therefore, $s$ and $L$ can be approximated  
$\rho_{\beta}(0)\frac{\overline{\sqrt{\xi}}}{\sqrt{\overline{\xi}}}\theta$ in the NND and NV expressions.

Within the same level of approximation we have 

\begin{equation}
4\int_{0}^{\infty}d\xi w(\xi)N_{\beta}^2\left(\sqrt{\frac{\xi}{\bar{\xi}}}\frac{\theta}{2}\right)
\simeq\left[\rho_{\beta}(0)\theta\right]^2=\left(\frac{\sqrt{\overline{\xi}}}{\overline{\sqrt{\xi}}}\right)^{2}L^2.
\end{equation}

Finally, if the disorder is defined by the distribution Eq. (\ref{Ts}) then we further have

\begin{equation}
\overline{\sqrt{\xi}}=\frac{\Gamma(\bar\xi + \frac{1}{2})}{\Gamma(\xi)}
\simeq\sqrt{\bar\xi}\exp\left(-\frac{1}{8\bar\xi}\right), \label{B5}
\end{equation}
where Stirling approximation has been used.

\section{Description of the empirical database}

The 467 Portuguese texts and 105 Chinese texts are downloaded from the Gutenburg project (www.gutenberg.org).  

The four Portuguese texts in the main texts are as follows: E. de Queiroz: Os Maias, 1888; E. de Queiroz: O Primo Basilio, 1878; C. C. Branco: A Filha do Arcediago, 1868; J. G. Rosa: Grande Sertao: Veredas, 1956.

The four Chinese texts in the main texts are as follows:  Xue Qin Cao: Hong Lou Meng (HLM), 18th century; Nai An Shi: Shui Hu Zhuan (SHZ), 14th century; Yao Lu: Ping Fan de Shi Jie (PFD), 1986; Mao San: Sa Ha La Sha Mo de Gu Shi (SHL), 1976. 

The 57 Protein sequence datasets are downloaded from www.uniprot.org and the 56 DNA sequence datasets are downloaded from www.ncbi.nlm.nih.gov/gene.

All the above analyzed datasets, the fitting codes and more empirical datasets fitting results are available from \cite{data}.


\begin{thebibliography}{99}
\bibitem{Poisson} To illustrate this point, we observe that if a spectrum is constructed by generating 
a sequence of points from the Wigner surmise, the NND of course will be Wigner but the NV is $L/4.$
\end{thebibliography}
\end{document}